\documentclass[11pt]{iopart}
\usepackage{graphicx}

\begin{document}
\title{Direct photons in d+Au and p+p collisions}

\author{M J Russcher for the STAR collaboration}

\address{Institute for Subatomic Physics, Utrecht University,\\
  3508 TA Utrecht, The Netherlands}
\ead{m.j.russcher@phys.uu.nl}

\begin{abstract}
Results are presented from an ongoing analysis of direct photon
production with the STAR experiment at RHIC. 
The direct photon measurement in d+Au collisions and 
the $\pi^0$ spectrum in p+p collisions are found to be 
in agreement with NLO pQCD calculations.
\end{abstract}


\section{Introduction}
\label{intro}
Direct photons, i.e. those not originating from hadron decays, 
are an interesting tool to study the quark-gluon plasma (QGP) created
in ultra-relativistic heavy-ion collisions \cite{PTphotons}. 

\textit{Thermal} direct photons are expected to be radiated by the electric 
charges in the QGP and the hadron gas, which is formed in a later stage 
of the expansion. A measurement of thermal photons can thus provide 
information on the temperature evolution of the system \cite{TRGphotons}. 

\textit{Prompt} direct photons originate from initial hard scatterings. 
They allow to verify scaling assumptions of initial parton densities and 
may be used to tag recoiling jets and study jet quenching
with a calibrated probe \cite{XNgammajet}. In addition their yield has 
to be known for studies of thermal photons, because they form a 
background to the latter.

It is expected that at RHIC thermal photons are most abundantly produced in Au+Au collisions,
which are beyond the scope of this paper. We investigate direct photon 
production in p+p and d+Au collisions in order to compare prompt 
photon measurements to pQCD predictions and study modifications in
cold nuclear matter. Both measurements will constitute a necessary 
reference to interpret the photon results obtained from Au+Au collisions.

\section{Analysis}
\label{sec:1}
The data presented here were taken with the STAR detector \cite{STAR}
in the 2003 d+Au and 2005 p+p run at RHIC, both at $\sqrt{s_{NN}}=200\,\mathrm{GeV}$.
The main detectors for this analysis are the Barrel Electromagnetic Calorimeter
(BEMC) \cite{BEMC}, the Barrel Shower Maximum Detector (BSMD) \cite{BEMC},
and the Time Projection Chamber (TPC) \cite{TPC}.
In order to enhance the raw particle yield at high transverse momentum ($p_T$)
an online trigger selected events with a high transverse energy deposition in
a single BEMC cell. More details on this direct photon analysis are given
in \cite{Russcher}.

This analysis aims to measure the direct photon yield by means of a
statistical subtraction of the hadronic decay background. The largest
contribution to this background comes from the decay 
$\pi^{0} \rightarrow \gamma\gamma$. It is therefore essential to measure
the pion spectrum with high precision. 

Inclusive photon candidates were identified by a clustering 
algorithm based on the energy measured in the BEMC and on the lateral shower 
profile measured in the BSMD. The latter is essential to resolve the two
$\pi^0$ decay photons at high $p_T$. To identify neutral clusters and decrease
hadronic background, a charged particle veto 
(CPV) is provided by rejecting clusters with a pointing TPC track.

\begin{figure}[h!]
\centering
  \scalebox{.45}{\includegraphics{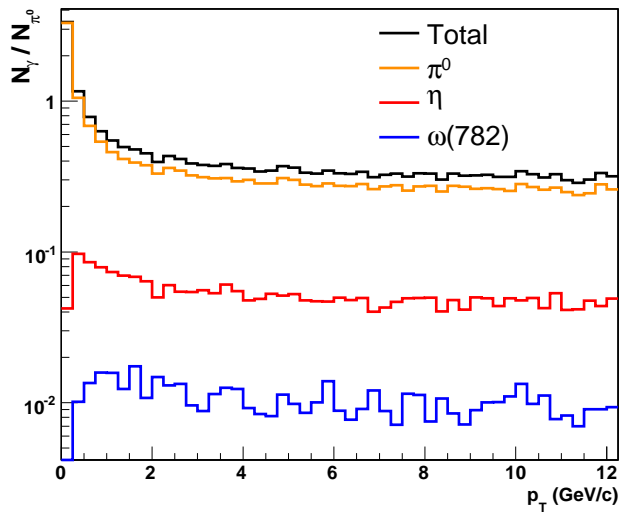}}
  \caption{The number of photons per input pion $N_{\gamma}/N_{\pi^0}$
    from hadronic decays versus $p_T$, obtained from a simulation 
    described in the text.}
\label{fig:bgphotons}
\end{figure}

The yield of decay photons has been determined from a Monte-Carlo
simulation which had a fit to the measured pion spectrum as input.
Since there are no STAR measurements available for $\eta$ and $\omega$(782)
yields, it was assumed that their transverse mass ($m_T$) spectra are the
same as that of the $\pi^0$ scaled by a constant which was taken
from literature \cite{PHetas,ISRomegas}. The result of the simulation
is shown in Figure \ref{fig:bgphotons}. Other possible contributions
to the decay background, such as $\eta'$ and $K^0_s$, were found to be negligible.

Showers from neutral hadrons and especially anti-neutrons can be misidentified
as photons. To correct for this contamination of the inclusive photon yield
a GEANT simulation of the detector was used which had the proton and 
anti-proton spectra as an input \cite{STprotons}. 
The data have been corrected for reconstruction and trigger efficiencies,
limited acceptance, photon conversions in the detector material
in front of the BEMC, and the inefficiency of the CPV.

\section{Results and outlook}
A convenient way of studying direct photons is by means of a double ratio 
\begin{equation}
  \label{eq:1}
  R = \frac{(\gamma/\pi^{0})_{measured}}{(\gamma/\pi^{0})_{decay}} 
  = 1 + \frac{\gamma_{direct}}{\gamma_{decay}}
\end{equation}
where the numerator is the point by point ratio of the measured inclusive photon
spectrum to the neutral pion spectrum and the denominator is the number
of simulated decay photons per input pion shown in Figure \ref{fig:bgphotons}.
The advantage of this ratio is that systematic uncertainties
which are common to neutral pion and inclusive photon detection will largely cancel. 

\begin{figure}[h]
  \centering
  \scalebox{.45}{\includegraphics{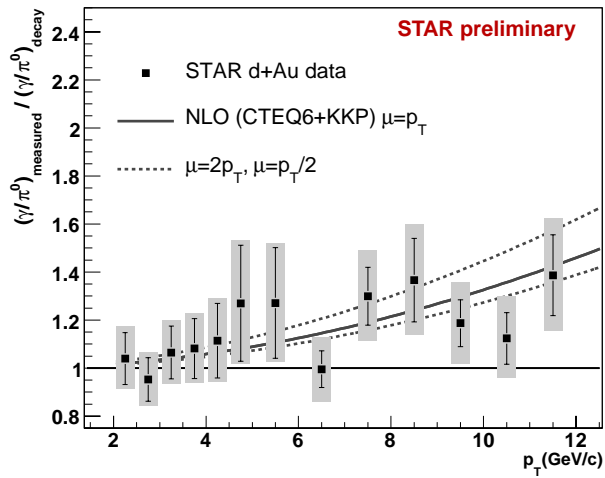}}
  \caption{Direct photon yields in  minimum bias $\sqrt{s_{NN}}=200\,\mathrm{GeV}$ d+Au
    collisions in terms of the double ratio $R=1+\gamma_{dir}/\gamma_{decay}$,
    see Equation \ref{eq:1}. The error bars (grey boxes) indicate the statistical (total) 
    error on the data points. The full line is a pQCD calculation 
    desribed in the text. The dashed lines show the sensitivity of 
    the calculation to the factorization scale.}
\label{fig:ratio_dau}
\end{figure}

Figure \ref{fig:ratio_dau} shows the $\gamma_{dir}$ measurement obtained from 
minimum bias d+Au collisions at $\sqrt{s_{NN}}=200\,\mathrm{GeV}$ in terms of $R$. 
The measurement is consistent with a pQCD calculation \cite{pQCDvogel} based
on the CTEQ6M parton density functions \cite{CTEQ6m}, the GRV parton-to-photon
\cite{GRV}, and the KKP parton-to-pion fragmentation functions \cite{KKP} as
shown for three different factorization scales by the curves
in Figure \ref{fig:ratio_dau}. The parton-to-photon fragmentation functions
are included since this analysis does not make use of an isolation cut for
the photon candidates.
The dominant systematic error on the measurement of the $\pi^0$ yield 
($\sim$30\%) is due to the uncertainty of the BEMC energy scale. 
This uncertainty largely cancels ($\leq$3\%) when calculating $R$ such 
that the $\pi^0$ yield extraction (5--10\%) and the statistical errors on the efficiency
correction (5--10\%) are the most important contribution to the error on $R$.

\begin{figure}
  \centering
  \scalebox{.5}{\includegraphics{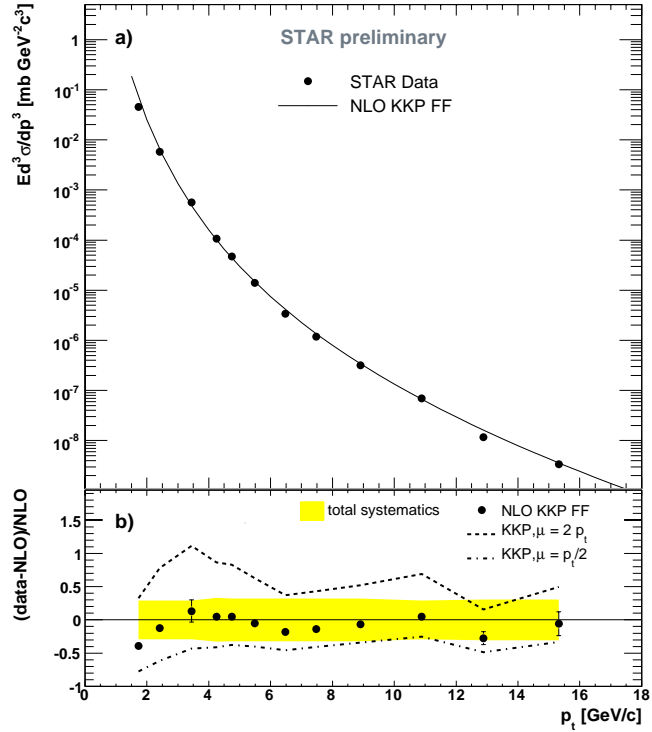}}
  \caption{The differential cross-section $Ed^3\sigma/dp^3$ for inclusive $\pi^0$
    production in p+p collisions at $\sqrt{s_{NN}}=200\,\mathrm{GeV}$ 
    as a function of $p_T$. The data are compared to a 
    pQCD calculation described in the text.}
  \label{fig:pions_pp}
\end{figure}

Figure \ref{fig:pions_pp} shows the $\pi^0$ cross-section in p+p collisions
as a function of $p_T$. Details of this analysis can be found in \cite{Simon}. 
The data is well described by a pQCD calculation
in combination with the KKP set of fragmentation functions \cite{KKP}
which is consistent with an earlier measurement at RHIC \cite{PHpions}.

The direct photon analysis of the 2005 p+p data set is in progress. 
The combined d+Au and p+p results will provide insight into nuclear 
effects and form a necessary baseline to study the properties 
of the QGP using direct photon measurements in Au+Au collisions.

\section{References}
\label{sec:references}

\end{document}